\documentclass[10pt]{article}  
\usepackage[margin=1in]{geometry}
\usepackage{graphicx}
\usepackage{url}
\usepackage[utf8]{inputenc}
\usepackage[english]{babel}
 
\usepackage{amsthm}
\newtheorem*{Remark}{Remark}
\title{Quantitative predictive modelling approaches to understanding rheumatoid
arthritis: A brief review}

\author{Fiona R Macfarlane$^{1}$, Mark AJ Chaplain $^{1}$ and Raluca Eftimie $^{2}$}
\date{$^{1}$ School of Mathematics and Statistics, University of St Andrews, UK;\\
$^{2}$ Department of Mathematics, University of Dundee, UK;}

\usepackage{latexsym}
\usepackage{amssymb,amsbsy,amsmath,amsfonts,amssymb,amscd,amsfonts,mathrsfs}
\usepackage{graphicx}
\usepackage{color}

\begin{document}
\maketitle
\begin{abstract}
Rheumatoid arthritis is a chronic autoimmune disease that is a major public health challenge. The~disease is characterised by inflammation of synovial joints and cartilage erosion, which lead to chronic pain, poor life quality and, in~some cases, mortality. Understanding the biological mechanisms behind the progression of the disease, as well as developing new methods for quantitative predictions of disease progression in the presence/absence of various therapies is important for the success of therapeutic approaches. 
The aim of this study is to review various quantitative predictive modelling approaches for understanding rheumatoid arthritis. To this end, we start by briefly discussing  the biology of this disease and some current treatment approaches, as well as emphasising some of the open problems in the field. Then, we review various mathematical mechanistic models derived to address some of these open problems. We~discuss models that investigate the biological mechanisms behind the progression of the disease, as well as pharmacokinetic and pharmacodynamic models for various drug therapies. Furthermore, we highlight models aimed at optimising the costs of the treatments while taking into consideration the evolution of the disease and potential complications.
\end{abstract}
 \section{Introduction}
Rheumatoid arthritis (RA), the~most common type of autoimmune arthritis~\cite{Merola2018_RheumatoidPsoriaticArthritis}, is a chronic autoimmune disease characterised by persistent inflammation of joints that leads to the damage of cartilage and, ultimately, bone within the joint~\cite{Calabresi2018_RAreview}. This~disease, which affects 1\% of the UK population~\cite{nrasstats}, impacts quality of life and even life expectancy (as RA can further lead to elevated risk of cardiovascular events~\cite{chimenti2015interplay,nhschoices}, reduced cognitive function in the brain, fibrotic disease in the lungs, osteoporosis, and~a greater risk of cancers~\cite{guo2018rheumatoid}). The~main symptoms of RA are pain, inflammation, swelling and stiffness of synovial joints. However, whole body symptoms of this disease can include tiredness, high temperature, weight loss and a loss of appetite~\cite{nhschoices}. There are various mechanisms involving genetic and environmental factors that contribute to RA pathogenesis, but the complex interactions between these mechanisms are not yet fully understood~\cite{Calabresi2018_RAreview}. For~example, it~is known that congenital predisposition is a risk factor for developing RA~\cite{Calabresi2018_RAreview}, with the heritability of RA being estimated at approximately 60\%~\cite{MacGregor2000_GeneticContributionRA}. The~role of genetic risks in RA development has been acknowledged since the 1980s, when the HLA-DRB1 (human leukocyte antigen DRB1 isotype) alleles were identified~\cite{Okada2018_GeneticsRAreview}. Advances in genome-wide association study technologies over the past 20 years have seen the identification of more than 100 genetic risk loci (i.e., chromosomal regions associated with the risk of disease)~\cite{Okada2018_GeneticsRAreview}. Nevertheless, despite this progress, there are still many open problems, such as the lack of conclusive identifications of causal genes or causal variants that might be responsible for the heterogeneity of RA~\cite{Okada2018_GeneticsRAreview,Weyand1998_HeterogeneityRA}. Genetic susceptibility to RA can also interact with some lifestyle factors, such as smoking~\mbox{\cite{Chang2014_SmokingRA,Castro2017,firestein2017immunopathogenesis}}, leading to an increase in the risk of developing the disease.

Other lifestyle and environmental factors that can increase the risk of RA include alcohol consumption, diet and exposure to occupational and atmospheric agents, such as silica dust and carbon-derived nanoparticles~\cite{Castro2017,chimenti2015interplay,guo2018rheumatoid,Calabresi2018_RAreview,Scrivo2017_SodiumRoleRA}. Moreover, some bacterial infections (e.g., \emph{Porphyromonas gingivalis}~\cite{Azzi2017_PeriodontalMicroRA}) and viral infections (e.g., Epstein-Barr virus~\cite{Balandraud2018_EpsteinBarrVirus_RA}) have been associated with RA development~\cite{PicchiantiDiamanti2018_InfectiousAgentsAR}. Despite these associations, which involve deregulated immune responses to bacterial and viral infections, to~date, there have been no conclusive causality studies on the role of such infections to RA. Irrespective of the mechanisms behind RA pathogenesis, the~disease is characterised by uncontrolled innate and adaptive immune responses that lead to auto-antigen presentation and aberrant production of pro-inflammatory cytokines~\cite{Calabresi2018_RAreview}. Given the heterogeneity of RA in terms of genetics, environmental interactions, serotype, clinical course and response to targeted therapeutic agents (discussed in more detail in the next section), the current view is that RA is not only one disease but a syndrome, which~is the result of different pathological pathways that lead to variable outcomes and phenotypes in individual~patients.

In the following section, Section~\ref{Bio_sec}, we summarise the different phases in the development of the disease, in~the context of autoimmunity and inflammation. We~additionally discuss how the key biological mechanisms are targeted in the context of RA treatment. We~note here that the~purpose of this work is to consider quantitative approaches to describe RA; therefore, the biological details provided are those required for understanding the reviewed modelling approaches, and~not an extensive discussion on the pathology of RA. We~refer the reader to \cite{guo2018rheumatoid,firestein2017immunopathogenesis} for more robust reviews of the biological mechanisms within RA. In Section~\ref{openq_sec}, we consider some of the open questions that remain in understanding RA development and treatment. We~then highlight mathematical modelling approaches that have previously been used to describe biological and therapeutic aspects of RA in Section~\ref{math_sec}. Finally, we conclude in Section~\ref{Sect:Conclusion} with a summary of this work and potential directions for future investigation in the context of mathematical modelling of RA.
 \section{Key Biology in RA}
 \label{Bio_sec}
In health, the~immune system is finely balanced including tight regulation of pro-inflammatory and anti-inflammatory mechanisms, whereas in RA, this balance of immunity is disrupted. The~progression of rheumatoid arthritis occurs over different phases that start with the development of autoimmune responses, followed by local inflammation within the joint and conclude with joint cartilage and bone destruction~\cite{chimenti2015interplay}. This~immune response is mediated by various cell types and chemicals within the joint space (i.e., chemokines and cytokines). We~discuss in more detail these different phases, while emphasising the roles of several key cytokines.

 \subsection{Disease Risk and Initiation}
The exact mechanisms that initiate the autoimmune response which characterises RA are not well understood. However, many risk factors have been identified which are thought to play a role in the initiation of the disease. For~example, the~presence of circulating antibodies and increasing concentrations of pro-inflammatory cytokines can characterise pre or early stages of RA \cite{firestein2017immunopathogenesis}. Notably, these factors can also be used as diagnostic markers, although generally, patients will not be diagnosed until RA is well established.

The first RA-associated antibody to be observed was rheumatoid factor (RF), an autoantibody directed against the FC region of immunoglobulin molecules~\cite{chimenti2015interplay}. Additionally, a key marker for subtypes of RA is the presence or absence of anti-citrullinated protein antibodies (ACPAs) \cite{firestein2017immunopathogenesis,van2018preventing}, which can be detected long before joint symptoms, e.g., pain and swelling. These ACPAs can be found in almost 67\% of RA patients and indicate a more aggressive form of RA that responds to immune cells and~treatments, in~a differing manner from the ACPA-negative form of the disease. The~presence or absence of these antibodies can be linked to genetic and environmental factors. Furthermore, these antigens can activate T cells that in turn help B cells produce more ACPA, leading to bone loss, inflammation and induction of pain in joints~\cite{chimenti2015interplay,guo2018rheumatoid}.

 In addition to RA associated antibodies, several studies have shown an increase in the level of particular cytokines, cytokine-related factors and chemokines prior to the onset of the disease. For~example, in~\cite{Kokkonen2010_CytokinesPredatesOnsetRA}, the authors observed an increase in IL-1$\beta$, TNF-$\alpha$, IL-6, IL-12 and GM-CSF levels in RF-positive and ACPA-positive individuals compared to control individuals. Therefore, it~was suggested that by combining the cytokine profile with the autoantibody status of RA patients, it~may help improve the early detection of the disease~\cite{Kokkonen2010_CytokinesPredatesOnsetRA,Chalan2016_ImmuneMarkers_earlyRA,Burska2014_Cytokines_BiomarkersRA}. Further methods of detection of RA include testing the erthrocyte sedimentation rate (ESR) and C-reactive protein (CRP) levels, which are correlated with severity of RA and their presence can indicate disease progression~\cite{heidari2011rheumatoid,nhschoices,van2018preventing}. 
 
Along with other risk factors mentioned in the introduction, such as heritability, smoking and viral risks, the~presence of these RA-associated antibodies and chemokines can increase the probability of the disease developing. Therefore, these mechanisms may play some role in initiating the autoimmune response that develops into established RA. 

Although less common, patients can have seronegative RA, where they exhibit symptoms of clinical RA without expressing the key markers, e.g., RF and/or ACPA. In these cases, it has been shown that HLA-DR4 can be used as a marker for disease severity~\cite{alarcon1982seronegative}. Patients who are seronegative may exhibit more severe inflammatory symptoms~\cite{nordberg2017patients}. Additionally, due to the increased severity of symptoms, seronegative patients generally receive more intensive treatment~\cite{smolen2017eular}.

 \subsection{Disease Progression}
The progression of RA is characterised by the infiltration of both adaptive and innate immune cells into the synovium of a joint, along with the expression of pro-inflammatory cytokines and chemokines, which lead to the inflammation of the synovial membrane. The~synovial membrane is a connective tissue made up of two layers, the~synovial lining and the synovial sublining. The~synovial lining is made up of macrophage-like synoviocytes and fibroblast-like synoviocytes (FLSs), whereas the synovial sublining is a soft connective tissue that controls smooth movement of joints~\cite{chimenti2015interplay}. Synovial cells, such as FLSs, are vital for normal joint function as they secrete the substances required for joint lubrication and movement \cite{firestein2017immunopathogenesis}.
 
The inflammation of the synovial membrane results in hyperplasia of the synovial lining~\mbox{\cite{Castro2017,fox2010cell,guo2018rheumatoid}}. This~synovial hyperplasia called the `pannus' is a thickening of the synovial lining caused by increased immune cell proliferation, invasion of immune cells from the circulation and local hypoxia driving angiogenesis. This~angiogenesis then promotes further infiltration of inflammatory cells and the production of pro-inflammatory cytokines. More specifically, macrophages can infiltrate the synovial membrane and produce NF-$\kappa$B, a transcription factor which promotes the production of pro- inflammatory cytokines, and~TNF-$\alpha$, a pro-inflammatory cytokine \cite{firestein2017immunopathogenesis}. Note, the~production of these pro-inflammatory cytokines can be enhanced in ACPA-positive RA, in~comparison to ACPA-negative RA. Normally, the~level of B cells is controlled by the balance between T regulatory cells (that inhibit B cells) and T helper cells (that promote B cells), however, in RA T helper cells increase the promotion of B cells, tipping this balancing process. Additionally, T helper cells can regulate macrophages by inducing their production of the pro-inflammatory cytokine TNF-$\alpha$~\cite{Castro2017}.

Synovial inflammation is partially dependent on migration of immune cells into the inflamed area combined with a lack of inflammatory cell death. T cell invasion into the inflamed area is enabled by mechanisms of leukocyte recruitment in the synovial vessels. The~leukocyte adhesion cascade requires coordination of rolling, adhesion and transmigration events. Specifically, the~leukocytes roll along the endothelium, become activated, adhere to endothelial cells and then migrate into the target tissue. This~T cell migration can become enhanced in RA, leading to a larger number of pro-inflammatory cells infiltrating the synovium~\cite{mellado2015t}. 

Furthermore, the~imbalance between inflammatory macrophages and anti-inflammatory macrophages is key in the formation of the pannus~\cite{guo2018rheumatoid}. Further cytokines, such as TNF-$\alpha$ and IL-15, produced by FLSs can also promote T cell migration, T cell activation and reduced FLS/T cell apoptosis in the inflamed area. Permeability and leakage of the vasculature may also lead to an influx of pro-inflammatory immune cells entering the joint. Angiogenesis additionally depends on the cytokine TNF-$\alpha$ and can promote invasion into the joint by inflammatory cells~\cite{fox2010cell}. Once the immune cells have initiated the formation of the pannus they secrete cytokines such as IFN-$\gamma$, IL-12, TNF-$\alpha$ and GM-CSF, which are all pro-inflammatory factors signalling further immune cells to enter the joint~\cite{chimenti2015interplay}.
 
The formation and growth of the pannus destroys the cartilage of the affected joint through adhesion and invasion processes~\cite{fox2010cell,chimenti2015interplay}. Cartilage is a connective tissue that consists of chondrocytes and a dense, highly organised, extra cellular matrix (ECM), which is produced by the chondrocytes. The~destruction of the ECM leads to further pro-inflammatory chemokines being released, stimulating FLS activity. Matrix metalloproteinases (MMPs) and tissue inhibitors of metalloproteinases (TIMPs) mediate cartilage destruction  in the RA setting~\cite{Castro2017,chimenti2015interplay,firestein2017immunopathogenesis,guo2018rheumatoid}. Once the cartilage is degraded, the bone can become exposed allowing for degradation of the bone.

In the later stages of the disease, bone erosion/loss is induced by cells within the bone called osteoclasts. Cytokines such as TNF-$\alpha$, IL-6, IL-1$\beta$ and IL-17 are all observed to exhibit pro-osteoclastogenic behaviour and can suppress bone formation through receptor activator of NF-$\kappa$B ligand (RANKL). These pro-inflammatory molecules are released by leukocytes, especially T helper cells, within the synovium leading to the promotion of bone erosion~\cite{Castro2017,guo2018rheumatoid,firestein2017immunopathogenesis}.

 \subsection{Treatment Approaches}\label{Sect:Treatment}
 \label{treatment}
A key factor in the effectiveness of therapeutic treatment of RA is how early the patient is diagnosed. In the last 10 years, the~classifications of early disease have been altered. Previously, any patient with disease duration of less than 2 years were termed `early disease' patients, however, more recently this has been reduced to those who have been diagnosed within 3 months~\cite{heidari2011rheumatoid,van2018preventing}. Recently, there has been investigation into preventative measures for RA,~\cite{van2018preventing}, although in this work, we focus on treatment of established RA. Currently, there is no known cure for RA, however, treatments can reduce disease progression, ease symptoms and prevent further joint damage~\cite{guo2018rheumatoid}. Patients may additionally go through `flare up' periods, where their symptoms become more acute for a short period of time before easing again. Furthermore, symptoms may become more prevalent after periods of inactivity, for~example, first thing in the morning. The~frequency and severity of these flare ups can be managed and reduced through treatment~\cite{nhschoices}. Although the damaging effects of RA on joints and bones cannot be reversed, a `remission' stage can be reached. Remission of RA can be classified in several different ways, such as, low inflammation indicated from blood tests, little or no joint swelling, little or no joint tenderness or less stiffness of joints. Remission of untreated RA occurs in 10\% of cases, however, relapse can also be common. With treatment, remission becomes much more likely and is correlated to how early the disease is diagnosed~\cite{heidari2011rheumatoid}. Due to the nature of the disease, most patients will continue to undergo long-term treatment and be monitored throughout the course of their treatment~\cite{nhschoices}. In~general, most treatment approaches work better in early forms of RA than in established cases of the disease~\cite{heidari2011rheumatoid}. 
 
There are several classes of RA treatment drugs including non-steroidal anti-inflammatory drugs (NSAIDs), steroids and disease-modifying anti-rheumatic drugs (DMARDs)~\cite{guo2018rheumatoid}. These drug based treatments can be used in conjunction with other drugs and with supportive treatments such as physiotherapy or occupational therapy. Additionally, often pain killers such as paracetamol or co-codamol are prescribed to reduce pain. Finally, in~severe cases, surgery may be an option. In the following paragraphs, we describe, in~further detail, some of the current drug types and approaches currently used in RA treatment. We~highlight here that the methods we describe are those that will be considered within later sections of this work. Therefore, we only consider a limited number of drugs and treatment options; for~a more extensive list we refer the reader to~\cite{nhschoices,heidari2011rheumatoid,sardar2016old}.

\paragraph*{\bf Non-steroidal anti-inflammatory drugs (NSAIDs)}
In the context of RA, NSAIDs are used to reduce pain and inflammation within joints. Ibuprofen, Naproxen, Diclofenac and COX-2 inhibitors are the most common NSAIDs prescribed to RA patients. Although an uncommon side effect, NSAIDs can reduce stomach lining which protects against stomach acids causing stomach problems such as internal bleeding. This~can be counteracted through the use of proton pump inhibitors (PPIs) that reduce the amount of stomach acid~\cite{nhschoices}.

\paragraph*{\bf Steroids}
Steroids can reduce joint pain, stiffness and inflammation and can be either taken in tablet form, and~injection into the joint or an injection into the muscle. Steroids are only used short-term, as long-term use can lead to serious side effects such as weight gain, osteoporosis, muscle weakness and thinning of the skin~\cite{nhschoices}.

\paragraph*{\bf Conventional Synthetic DMARDs}
Conventional DMARDs can be used to reduce disease activity and reduce or delay joint deformity in RA patients~\cite{guo2018rheumatoid}. Furthermore, DMARDs can target and block the inflammatory chemicals within RA, however, they may take some time (e.g., months) to become effective~\cite{nhschoices}.
\begin{itemize}
\item \label{MTX}\textbf{Methotrexate (MTX)} can induce downregulation of MMPs and can be used to modify the patient's cytokine profile. It is administered weekly at a low dose, and~requires regular monitoring, through blood tests, to assess the immunosuppressive and hepatotoxic effects of the drug~\cite{guo2018rheumatoid}. MTX is generally the first drug prescribed to treat RA in the UK and has been shown to reduce pain, joint damage and other symptoms within a short time period. Common side effects can include sickness, loss of appetite and diarrhoea. However, more severe effects, such as kidney, lung and liver problems, can be experienced. Furthermore, there can be a high discontinuation (e.g., 16\% of patients) of the drug due to adverse effects~\cite{mtxref}.

\end{itemize}

\paragraph*{\bf Biological DMARDs}
Biological DMARDs are generally used in conjunction with conventional DMARDs, or~if conventional DMARDs are ineffective. They are usually administered via injection and aim to target pro-inflammatory cytokine pathways~\cite{nhschoices}. Biological DMARDs are considered safe in regard to their risk of infection and have a strong benefit-to-risk profile~\cite{singh2015risk}. Furthermore, they can be used safely in conjunction with other biological DMARDs or with conventional synthetic DMARDs (e.g., MTX). The~combination of MTX with these biological DMARDs can improve the efficacy of MTX alone~\cite{nam2017efficacy}. We~describe some commonly used biological DMARDs below:
\begin{itemize}
\item \label{IFX} \textbf{Infliximab (IFX)} is a TNF-$\alpha$ inhibitor (TNFi) which reduces the thickness of the synovial layer and has also been shown to lead to a decrease in the levels of IL-6~\cite{guo2018rheumatoid}.

\item \textbf{Etanercept} is another TNFi, with a similar toxicity profile to IFX~\cite{guo2018rheumatoid}.

\item \textbf{Adalimumab} is a TNFi which is considered safe to use, with minor side effects~\cite{adalimumabref}.

\item\textbf{Tocilizumab (TCZ)} targets and inhibits the IL-6 receptor which is produced by B cells, FLSs and monocytes. Patients, generally, exhibit a positive response to the drug, however, increased cholesterol and increased (minor) adverse effects are common side effects~\cite{tczref}.
\end{itemize}

\paragraph*{\bf Surgery}
In severe cases of joint damage, surgery may be required. For~example, carpal tunnel release is where a ligament is cut in the hand to relieve pressure on the nerves or, alternatively, tendons may be cut to treat abnormal bending. In some cases, arthroscopy is used to remove inflamed tissue around the affected joints. In the case of total joint destruction, joint replacement can be used. As most prosthetics have a lifespan of 10--20 years, surgery may have to be repeated~\cite{nhschoices}.

 \section{Open Questions in Rheumatoid Arthritis and the Role of Mathematical Modelling}
 \label{openq_sec}
 
As we have discussed in the previous section, in~recent years, there has been increased understanding of the mechanisms within RA progression and the development of successful RA therapies. However, there remains several unanswered questions and areas for further investigation within the area of RA research. These include the~development of earlier diagnostic techniques, understanding fully the mechanisms that initiate RA and prediction of the side effects of treatment~approaches. 

A key factor in the success of RA treatment is the diagnosis time-frame, with many studies highlighting the importance of early detection and referral on the overall improvement of patient outcomes~\cite{Mallen2018_EnhanceEarlyDiagnosisRA,Fleischmann1999_EarlyDiagnosisRA,daMota2015_ImportanceEarlyDiagnosisRA,Cho2017_FactorsTimeToDiagnosisRA}. The~challenges in identifying this disease at an early stage are related to: (i) the very low incidence of RA (i.e., 15--40 cases per 100,000 adult patients~\cite{Mallen2018_EnhanceEarlyDiagnosisRA}) which reduces the likelihood that a general practitioner will identify it in its early stages, and~(ii) the presence of non-specific symptoms in the early phases of the disease~\cite{Mallen2018_EnhanceEarlyDiagnosisRA}. For~example, many people may exhibit joint pain and stiffness, especially in cases where the joints are under excessive strain (e.g., overweight patients or athletes). However, in~these situations, joint pain is likely to be a symptom of natural joint damage, or~in more severe cases the development of osteoarthritis, and~not the autoimmune-linked RA~\cite{nhsosteo}. Therefore, improved methods of detecting risk factors of RA and the disease symptoms earlier should be a focus for further research. 
 
Although the mechanisms of inflammation that allow for progression of rheumatoid arthritis are well understood, the~mechanisms which tip the balance of immunity to initiate inflammation are still not fully understood. Many therapeutic approaches, in~RA and other autoimmune diseases, aim to restore the balance of immunity~\cite{flores2008restoring,smilek2014restoring}. Therefore, by understanding the key mechanisms involved in this inbalance of immunity, the~development of new restorative treatments can be achieved.
 
Treatment of RA may not be successful for all patients, due to the heterogeneous nature of the disease~\cite{Weyand1998_HeterogeneityRA}. This~can result in `Difficult-to-treat' RA, which can be defined as the persistence of symptoms despite treatment with conventional DMARDs and at least two biological DMARDs \cite{de2017difficult}. Various factors may contribute to continued disease progression under treatment, such as adverse drug reactions, pharmacogenetics, lifestyle factors and genetic factors. In relation to drug ineffectiveness, patients may be prescribed multiple forms of RA treatments at once, or~switch treatments over the course of their disease. As mentioned previously, some of the drugs used to treat RA can have mild to severe side effects. Moreover, some of these side effects may depend on which other treatments the patient is currently taking, or~has undergone in the past. For~example, the~risk of developing some infections can be higher when taking biologic DMARDs than conventional DMARDs~\cite{singh2015risk}. Furthermore, the~risk of mild or severe infections of patients taking TCZ can be increased if they have been exposed to more than three other DMARDs, especially MTX~\cite{campbell2010risk}, or~have had a long disease duration~\cite{lang2011risk}. These infections can range in severity, however, some can be life threatening if not treated quickly, e.g., sepsis and pneumonia. Therefore, greater understanding of the interplay between treatment types and the safety of treatment dosage is required to improve treatment success.

To address some of these open questions regarding the complex mechanisms of RA pathogenesis, the~introduction of new methods to improve early detection of the disease, or~the outcomes of different (combined) treatment approaches, the~last 20--30 years have seen the development of various types of mathematical models~\cite{Alder2014_ComputerBasedDiagnosisRA_Review,baker2013mathematical,helliwell2000joint,Witten2000_OneODE_ProgressionRA,rao2016mathematical,Levi2013_PKPDmodel_Tocilizumab_RA,Kobelt2005_CostEffectivenessRA,Kimura2014_PharmacokineticPharmacodynamicRA}. In the following section, we review some of these models. To this end, we focus only on mechanistic models for disease progression and treatment, while ignoring statistical approaches~\cite{Anderson1984_EstimateDiseaseActivityRA,Breedveld2005_StatisticalModelRadiographicDamage,Capela2014_StatsModel_DiseaseActivityScore,Terao2019_RAsubsetsGeneticsStatsModel}, or~machine learning approaches~\cite{LezcanoValverde2017_MachineLeaningRA,Norgeot2019_DeepLearningRAmodel,Kim2019_MachineLearningRA}.

We should mention here that in the mathematical literature, there is a very large variety of models that focus on autoimmune diseases, and~the immune mechanisms triggering and controlling these diseases; see, for~example,~\cite{Alexander2010_SelfToleranceAutoimmunityModel,Arazi2010_ModelAutoimmuneInflammation,Blyuss2012_ActivationThresholdAutoimmunity,Delitala2013_MathModelAutoimmuneResp,Fatehi2018_StochasticAutoimmuneDynamic,Fatehi2018_CytokineDelaysAutoimmunity,Iwami2007_DynamicsAutoimmuneDiseases,MachadoRamos2019_KineticModelAutoimmune,Rapin2011_BistabilityAutoimmuneDiseases,Smirnova1975_MathModelAutoimmunity} and the references therein. However, in~this study, we focus exclusively on those mathematical models derived specifically for rheumatoid arthritis, and~ignore the more general models (even if these models could also be applied to RA). 

 \section{Mathematical Modelling Approaches}
\label{math_sec}

\begin{figure}
\centering
\includegraphics[width=4.7in]{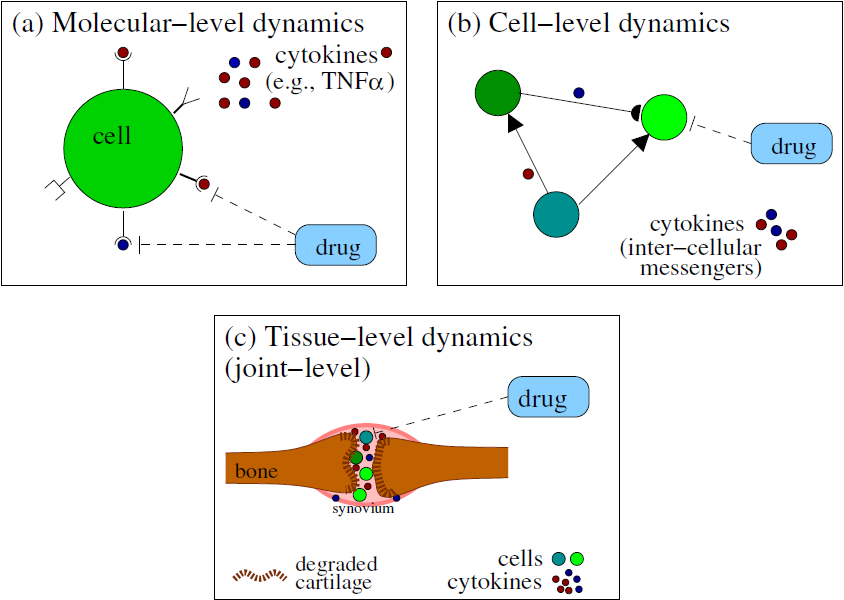}
\caption{Depiction of the spatial scales at which mathematical models can investigate RA dynamics: (\textbf{a}) the molecular scale; (\textbf{b}) the cell scale; (\textbf{c}) the tissue scale.}
\label{Fig_Multiscale}
\end{figure}

\begin{figure}
\centering
\includegraphics[width=1.9in]{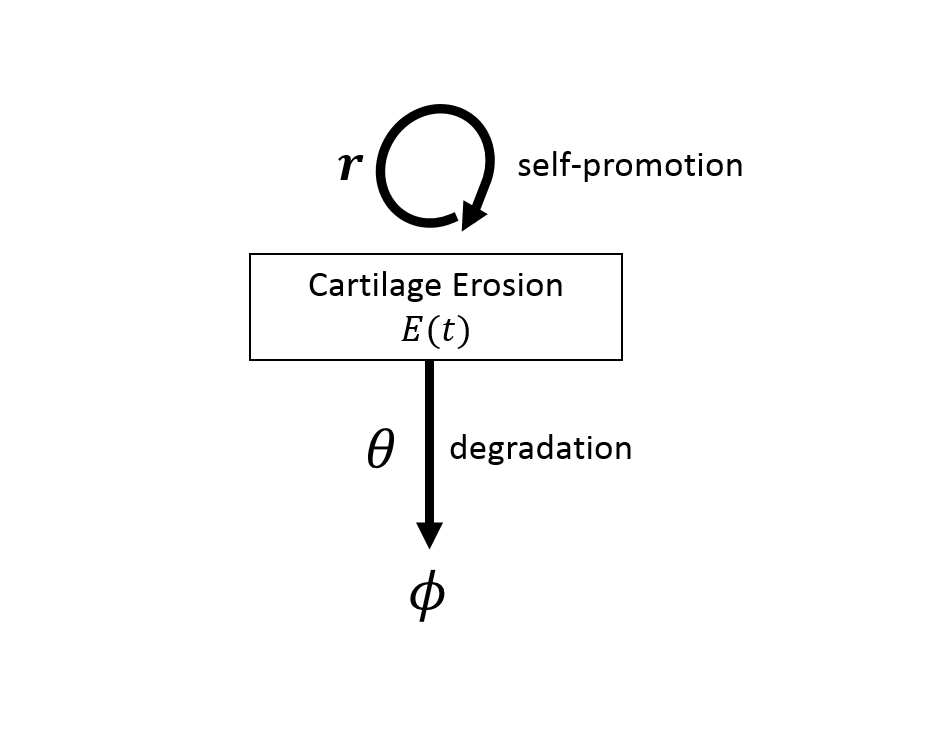}
\caption{Schematic description of the logistic growth/decay for cartilage erosion introduced in~\cite{Witten2000_OneODE_ProgressionRA}. The~model considers only one variable, the~cartilage erosion over time $t$, which is denoted by $E(t)$ in Equation~(\ref{LogisticEq}). Here $\phi$ represents cartilage removed from the system.}
\label{Fig_logistic}
\end{figure}

In this section, we summarise some of the mathematical modelling approaches derived over the past 2--3 decades to understand the pathogenesis of rheumatoid arthritis, as well as to model various treatments that could improve the clinical manifestation of the disease. We~emphasise that the models summarised in this section investigate RA dynamics at multiple spatial scales: from binding/unbinding of cytokines to cell receptors (at molecular level), which trigger a cascade of reactions that culminate with inflammatory processes, to cell--cell interactions via different cytokines (at~cell level), and~the degradation of the cartilage (at joint level); see also Figure~\ref{Fig_Multiscale}. As~a quick reference to the reader, we include in Table~\ref{Table_Cytokines} a summary of the cytokines that are considered within the mathematical models we review, and their key biological roles. In Table~\ref{TableModels}, we summarise the key methods generally used in mathematical modelling of RA, along with their advantages and disadvantages.
\begin{table}[!h]
\caption{{{Cytokines considered within mathematical models of RA and their biological role.}}}
\centering
\begin{tabular}{ll}
\hline
\textbf{Cytokine}		& \textbf{Role}\\
\hline
GM-CSF				& Promotes inflammation~\cite{Ridgley2019_DominantCytokinesRA}\\ 
					& Activates macrophages, neutrophils~\cite{McInnes2016_CytokinesRA}\\
					& \\
IFN-$\gamma$			& Increases antigen presentation~\cite{Castro2017} \\
					& Activates macrophages~\cite{Castro2017} \\
					& Increases chemokine secretion~\cite{Castro2017}\\
					& \\
					
IL-1$\beta$					& Induces osteoclastogenesis~\cite{Castro2017,heidari2011rheumatoid} \\	
&\\				
IL-6					& Activates leukocytes and osteoclasts~\cite{Castro2017,heidari2011rheumatoid,firestein2017immunopathogenesis} \\
 					& Stimulates antibody production~\cite{Castro2017} \\
					& Promotes pannus formation~\cite{guo2018rheumatoid}	\\
					& \\
IL-12					& Involved in the plasticity of Th17 (subset of T helper) cells~\cite{Ridgley2019_DominantCytokinesRA}\\					
					&\\
IL-15					& Promotes T cell migration~\cite{fox2010cell} \\												& \\
IL-17					& Induces production of inflammatory cytokines~\cite{Castro2017} \\
					& Activates innate immune cells~\cite{Castro2017} \\
 					& Induces osteoclastogenesis~\cite{Castro2017,heidari2011rheumatoid}\\
 					& Stimulates neutrophil recruitment~\cite{Castro2017}\\
					& \\
RANKL 				& {{Promotes bone erosion}}~\cite{Castro2017,heidari2011rheumatoid,firestein2017immunopathogenesis}\\
&\\
TNF-$\alpha$			& Activates leukocytes, FLSs, endothelial cells and osteoclasts~\cite{Castro2017}
\\
					& Induces production of inflammatory cytokines~\cite{Castro2017}\\
					& Enhances MMP production~\cite{Castro2017,heidari2011rheumatoid}\\
					& Suppresses T regulatory cells~\cite{Castro2017}\\
					& Promotes T cell migration~\cite{fox2010cell}\\
					& Activates the RANKL pathway~\cite{heidari2011rheumatoid}\\
					& Promotes osteoclastogenesis~\cite{Castro2017,heidari2011rheumatoid,firestein2017immunopathogenesis}\\
					& Promotes angiogenesis~\cite{fox2010cell}\\
					& \\
\hline
\end{tabular}
\label{Table_Cytokines}
\end{table} 
\begin{table}[!hp]
\caption{{Summary of the main types of mathematical models used to describe the evolution and possible treatment approaches for RA.} }
\centering
\resizebox{\textwidth}{!}{\scriptsize
\begin{tabular}{p{3.5cm}p{10.5cm}}
\hline
{{\textbf{ Math Models}}}		& {{\textbf{ Description/Advantages/Disadvantages}}}\\
\hline
Ordinary Differential Equations (ODEs)	&	{Deterministic mathematical equations that describe the time evolution of a variable of interest; e.g., the density of immune cells involved in RA, the density of chondrocytes in the cartilage, the concentration of some cytokines, or the concentration of a therapeutic drug. These are the most common models used to describe the evolution of RA~\cite{Witten2000_OneODE_ProgressionRA,baker2013mathematical,jit2004tnf,matteucci2019solution,odisharia2017mathematical,rao2016mathematical,rullmann2005systems,Kimura2014_PharmacokineticPharmacodynamicRA}.} \\
&{ \emph{Advantages:} These types of models require shorter simulation time, and are easily parametrised using experimental lab data or clinical patients data (because the large majority of collected data describes the temporal changes in some variable of interest; e.g., levels of pro-inflammatory cytokines). These models can also be investigated analytically; e.g., their long-term dynamics can be studied via the identification of possible steady states and their stability~\cite{baker2013mathematical}. }\\
&{ \emph{Disadvantages:} These types of models cannot really capture the mechanisms behind the spatial degradation of the articular cartilage. Also, being deterministic, these models cannot capture the variability in the cytokine levels between different patients~\cite{Burska2014_Cytokines_BiomarkersRA}. This variability in the cytokine level can impact also the variability in the evolution of the disease. }\\
&\\
Partial Differential Equations (PDEs)	&  {Deterministic mathematical equations that describe the space and time evolution of a variable of interest (e.g., the density of immune cells involved in RA, the density of chondrocytes in the cartilage, the concentration of some cytokines, or the concentration of a therapeutic drug~\cite{moise2019rheumatoid}). They can also describe age-related aspects of the immune responses; however this approach is not very common in the context of RA dynamics. }\\
& { \emph{Advantages:} These models can be used to test various hypotheses on the spatial dynamics of the components of immune system involved in RA evolution (e.g., spatial spread of cytokines, immune infiltration of the joint, etc.), and how can these components affect the spatial erosion of the cartilage. They can also be used to investigate the spread of the therapeutic drug into the affected tissue. }\\
& { \emph{Disadvantages:} The numerical simulations of these models are more complex (compared to the simulations of ODEs). It is also more difficult to parametrise these types of mathematical models using patients data. As with the ODEs, since these models are deterministic, they cannot capture the variability in the cytokine levels between different patients~\cite{Burska2014_Cytokines_BiomarkersRA}. Due to the complexity of these models (coupled sometimes with a large number of variables modelled) it can be more difficult to investigate these models analytically. }\\
&\\
Stochastic/Probabilistic models	&{ Mathematical and computational models where the interactions between the different components of the system, or the  transitions between different states of these components are probabilistic. Such models have been mainly applied in the context of treatment decisions~\cite{Scholz2014_ModelRA}. Very few models have been used to describe the probability of RA occurrence. } \\

& { \emph{Advantages:} These models can reproduce more accurately the randomness of certain aspects in the evolution of RA (e.g., transition between different states as a results of different levels of pro-inflammatory cytokines, etc.). }\\

& { \emph{Disadvantages:} Due to slightly more complex numerical simulations, until now these models have been used to describe the temporal dynamics of different variables involved in the evolution of RA. As far as we know, probabilistic approaches have not been combined yet with spatio-temporal models to investigate the variability in the spatial dynamics of different pro-inflammatory and/or anti-inflammatory immune responses. Since these are mostly computational models, analytical investigations are inexistent.} \\	
\hline
\end{tabular}}
\label{TableModels}
\end{table} 
We start in Section~\ref{Sect:ODEs} by reviewing some simple and more complex, ordinary differential equation (ODE) models describing the evolution of RA, as well as the dynamics of drugs used for the RA treatment. In Section~\ref{Sect:PDEs}, we review a complex partial differential equation (PDE) model for the spatial movement of immune cells and chondrocytes in the various compartments of the joint (see Figure~\ref{Fig_Multiscale}c), as well as the diffusion of cytokines and drug molecules. In Section~\ref{Sect:Stochastic}, we discuss non-deterministic models, i.e., models that consider stochastic effects within RA. In Section~\ref{Sect:CostEffectiveness} we briefly discuss some probabilistic models used in the decision-making processes related to various RA treatment approaches and their overall costs. Finally, in~Section~\ref{Sect:Data}, we discuss parameter estimation and the availability of data, in~the context of mathematical modelling of RA. Because some models discussed below are described by a very large number of equations (e.g., 34 coupled differential equations in~\cite{rao2016mathematical}, or~17 coupled equations in~\cite{moise2019rheumatoid}), we decided not to list all these equations here. However, for~the models presented in more detail, we show diagrams with schematic descriptions of the interactions modelled by those equations. To help the reader, in~the next section we show how to translate the interactions encoded by these diagrams into mathematical equations, for~two simple cases: (i) a one-equation model introduced in~\cite{Witten2000_OneODE_ProgressionRA} to describe the time-evolution of cartilage erosion (see Figure~\ref{Fig_logistic}), and~(ii)~a two-equation model introduced in~\cite{baker2013mathematical} to describe the time-evolution of pro-inflammatory and anti-inflammatory cytokines, as a results of positive and negative feedback interactions between these cytokines; see Figure~\ref{Baker}. The~approaches detailed in these two examples can be easily extended to more complicated single-scale and multi-scale interactions, as those depicted by the diagrams in Figures~\ref{jit}--\ref{Moise}~below.

\unskip

 \subsection{Deterministic Models for Disease Progression and Treatment: ODEs}\label{Sect:ODEs}
Ordinary differential equations (ODEs), and~systems of ODEs, have been used over the past two decades to describe the evolution of RA in the presence and absence of therapies. Some of the simplest ODEs in the literature, and~the solutions of such simple ODEs, were used more than two decades ago to describe the increase in the amount of joint erosion as seen following radiographic examinations~\cite{FuchsPincus1992_RadiographicDamageRA,Salaffi1994_ProgressRAErosionNonlinModel,Sharp1994_CurveFittingModelsRAdamage,Graudal1998_RadiographicProgressRA}, thus these studies focused on tissue-level dynamics. However, since these equations described a linear or exponential growth in the erosion level, they were considered unrealistic (due to the fixed number of joints and cartilages available for erosion). To address this issue,~\cite{Witten2000_OneODE_ProgressionRA} proposed a logistic-type growth ODE model for the time-evolution of the joint erosion level, which could better predict all stages of the disease (i.e., a slow initial growth in joint erosion, followed by a linear-like growth, and~finally a slow-down in erosion). This~ODE model (for the time-evolution of the erosion variable $E(t)$) has the general form
\begin{equation}
\underbrace{\frac{dE(t)}{dt}}_{\text{time changes in}\; E(t)}=\underbrace{rE(t)}_{\text{increase in the erosion at rate} \; r}-\underbrace{\theta E(t)^{2}}_{\text{decrease in the erosion at rate }\;\theta}
\label{LogisticEq}
\end{equation}

The authors used the sigmoidal solution of this logistic-growth model to calculate the time-intervals associated with the different stages in RA progression: the ``pathology'' stage, the~``impairment'' stage, the~``functional limitation'' stage, and~the ``functional disability'' stage.

\begin{figure}
\centering
\includegraphics[width=.65\textwidth]{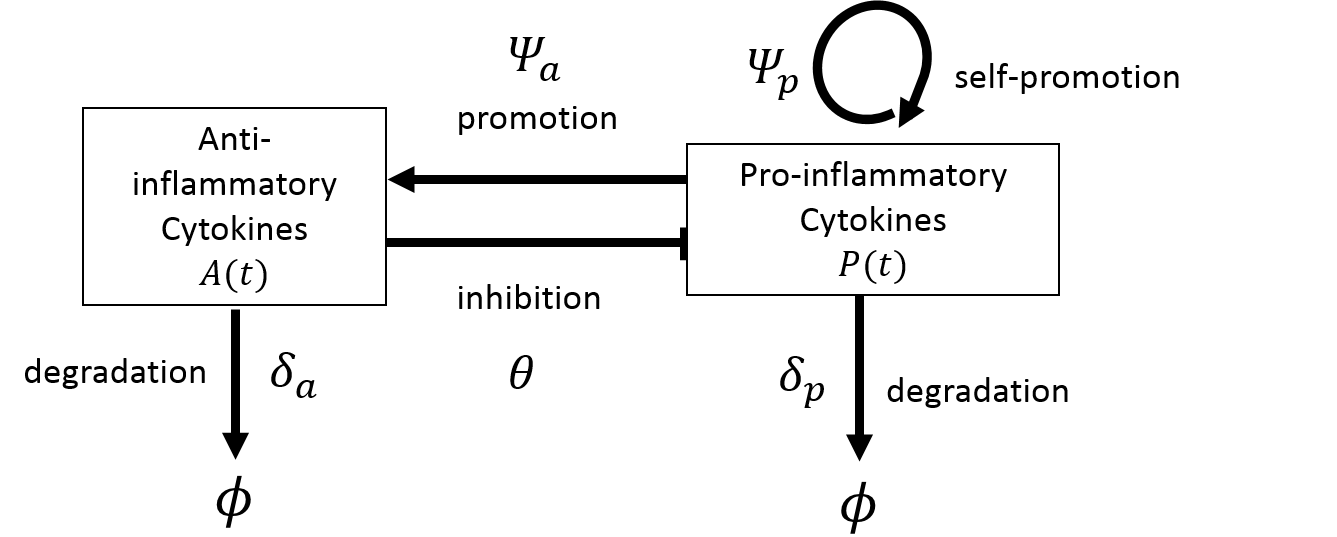}
\caption{Schematic describing the chemicals and processes included in model described in~\cite{baker2013mathematical}. The~models considers two variables, the~pro-inflammatory cytokines and anti-inflammatory cytokines, and~their interactions with each other. Note that, here $\phi$ represents objects removed from the system. The~interactions depicted in this diagram are described by Equation~(\ref{model:Baker}), where $A(t)$ represents the density of anti-inflammatory cytokines at time $t$, and~$P(t)$ represents the density of pro-inflammatory cytokines at time $t$. These two variables are shown in the two rectangles appearing in the figure above. The~arrows between theses two rectangles correspond to the interactions described by the terms in the right-hand side of Equation~(\ref{model:Baker}).}
\label{Baker}
\end{figure}

In the context of mechanistic approaches for the evolution of RA, Baker~et~al.~\cite{baker2013mathematical} developed a simple system of 2 ODEs to describe RA progression as determined by the interactions between pro-inflammatory and anti-inflammatory cytokines; see Figure~\ref{Baker}. The~authors completely neglected any variability in cell behaviour and focused on the dynamics of the cytokines produced by some generic cells in the synovium. The~model in~\cite{baker2013mathematical} incorporated the assumptions that (i)~both pro-inflammatory and anti-inflammatory cytokines decay naturally, (ii) the pro-inflammatory cytokines can stimulate the production of more pro-inflammatory cytokines and also the production of anti-inflammatory cytokines, and~(ii) the anti-inflammatory cytokines can reduce the production of pro-inflammatory~cytokines.

These assumptions, depicted schematically in Figure~\ref{Baker}, can be described mathematically by the following two coupled equations (where $A(t)$ denotes the concentration of anti-inflammatory cytokines at time $t$, and~$P(t)$ denotes the concentration of pro-inflammatory cytokines at time $t$):
\begin{subequations}
\label{model:Baker}
\begin{align}
\underbrace{\frac{d P(t)}{dt}}_{\text{time-change in}\; P(t)}=&\underbrace{-\delta_{p} P(t)}_{\text{degradation of}\; P(t) \; \text{at rate} \; \delta_{p}}+\underbrace{\theta(A(t))}_{\text{inhibition of}\; P(t)\; \text{by}\; A(t) }\boldsymbol{\cdot} \underbrace{\Psi_{p}(P(t))}_{\text{self-promotion of}\; P(t) },\\
\underbrace{\frac{d A(t)}{dt}}_{\text{time-change of }\; A(t)}=&\underbrace{-\delta_{a} A(t)}_{\text{degradation of}\; A(t)\; \text{at rate}\; \delta_{a}}+\underbrace{\Psi_{a}(P(t))}_{\text{promotion of}\; A(t)\; \text{by } P(t)}.
\end{align}
\end{subequations} 
The functions $\theta(A)$, $\Psi_{a}(P)$ and $\Psi_{p}(P)$ can take different forms, to phenomenologically describe the experimental observations. For~example, in~\cite{baker2013mathematical} the authors suggested that the production of anti- inflammatory and pro-inflammatory cytokines in the presence of $P(t)$ should have some maximum rate, as well as the down-regulation of $P(t)$ in the presence of $A(t)$. Thus they give some possible choices for these terms: $\theta(A(t))=c_{3}c_{4}^{2}/(c_{4}^{2}+A(t)^{2})$, $\Psi_{p}(P(t))=c_{0}+c_{1}P(t)^{2}/(c_{2}^{2}+P(t)^{2})$ and $\Psi_{a}(P(t))=c_{5}P(t)^{2}/(c_{6}^{2}+P(t)^{2})$, where $c_{0}$, $c_{1}$, $c_{2}$, $c_{3}$, $c_{4}$, $c_{5}$ and $c_{6}$ represent some non-negative constant parameters. The~authors identified the steady states for this simple model and investigated their stability. In addition, they used numerical bifurcation approaches to identify monostable and bistable behaviours. They also investigated numerically the effects of anti-inflammatory treatments, although these treatments were not specific to particular cytokines.

\begin{Remark}
Before we move on to discuss more complicated mathematical models that correspond to the complex diagrams shown in Figures~\ref{jit}--\ref{Moise} below, we need to emphasise that those diagrams can be translated into mathematical equations the same way as for the two cases described by Equations~(\ref{LogisticEq})--(\ref{model:Baker}). However, since the models discussed below incorporate more complex interactions and a larger number of variables, it~would be too tedious to write-down all corresponding equations. For~this reason, in~the following we present only the schematic diagrams depicting the biological interactions included in the models discussed here, and~for the detailed mathematical equations we refer the reader to the original papers.
\end{Remark}

For a more in-depth understanding of the specific mechanisms involved in RA progression more complex mathematical models can be developed. To exemplify this aspect, we start by mentioning the 4-ODE model developed by Jit~et~al.~\cite{jit2004tnf} to investigate the role of TNF-$\alpha$ in RA. The~authors focused on the molecular-level dynamics of TNF-$\alpha$, and~considered the formation of ligand-receptor complexes, antigen-antibody complexes and the binding process of TNF-$\alpha$ to cell surface receptors; see Figure~\ref{jit}. The~aim of the study in~\cite{jit2004tnf} was to predict the short-term and long-term benefits of various therapeutic approaches that inhibit TNF-$\alpha$, such as sTNFR2, etanercept and infliximab (see Section~\ref{treatment} for a more detailed discussion of the TNF-$\alpha$ inhibitors, etanercept and infliximab). The~authors first identified the equilibrium states for this model (e.g., a ``normal'' state with zero TNF-$\alpha$ production, and~a ``pathological'' state with non-zero TNF-$\alpha$), and~then varied certain parameters to replicate different treatment situations. Note that, in~this study the parameters were estimated based on various experimental data. Furthermore, Matteucci~et~al.~\cite{matteucci2019solution} obtained a general analytical solution for the ODE system developed by Jit~et~al.~\cite{jit2004tnf}, and~then used this solution to further evaluate the impact of the TNF-$\alpha$ inhibiting drugs, where different initial conditions are considered.
\begin{figure}
\centering
\includegraphics[width=0.8\textwidth]{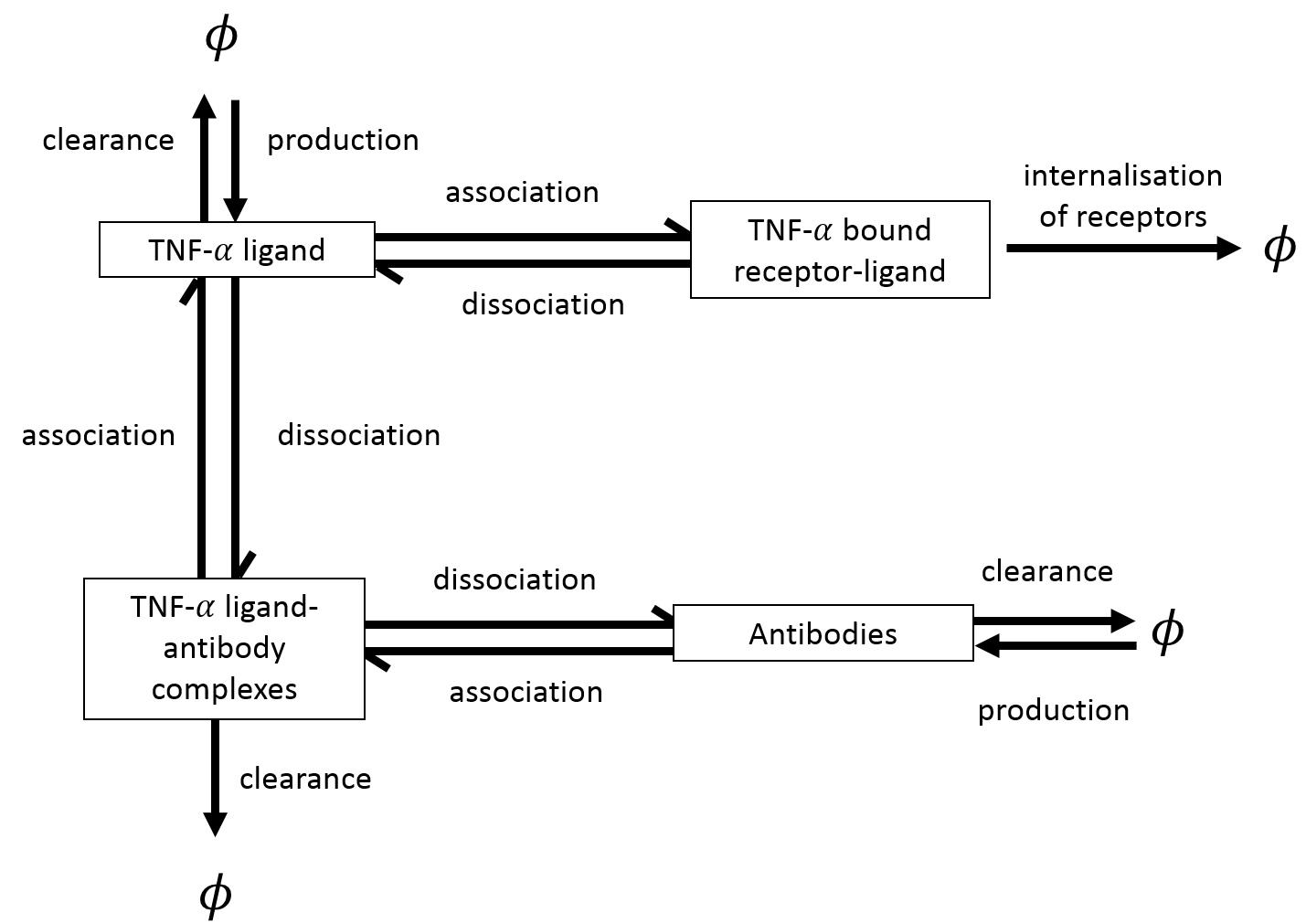}
\caption{ Schematic describing the chemicals, receptors and processes included in model described in~\cite{jit2004tnf}. The~models considers four variables; antibodies, TNF-$\alpha$, bound TNF-$\alpha$ and antibody-TNF-$\alpha$ complexes, along with their interactions with each other. Note that, here $\phi$ represents objects outwith or removed from the system.}
\label{jit}
\end{figure}

An example of a slightly more complex ODE model for RA progression can be found in Odisharia~et~al.~\cite{odisharia2017mathematical}, where the authors developed a system of 5 ODEs that was used to describe the interactions between B cells, T cells (i.e., T helper and T regulatory cells) and cells that form the cartilage, in~the presence of a drug, tocilizumab (TCZ) (which we described in Section~\ref{treatment}); see also Figure~\ref{Odishara}. The~model incorporated the assumptions that (i) T helper cells stimulate B cell activity, (ii)~T regulatory cells inhibit B cell activity, (iii) the drug TCZ blocks the growth of T helper cells and transforms them into T regulatory cells, (iv) B cells that reach levels higher than their normal values destroy the cartilage. The~authors investigated numerically the dynamics of these different cells, including the degradation of the cartilage, as various model parameters are varied. The~15~parameters within the model could be potentially calibrated from blood clinical analysis of patients, but the authors did not seem to use actual patients data for their simulation results. Furthermore, the~authors did not explicitly consider any side effects of the drug.
 \begin{figure}
\centering
\includegraphics[width=0.8\textwidth]{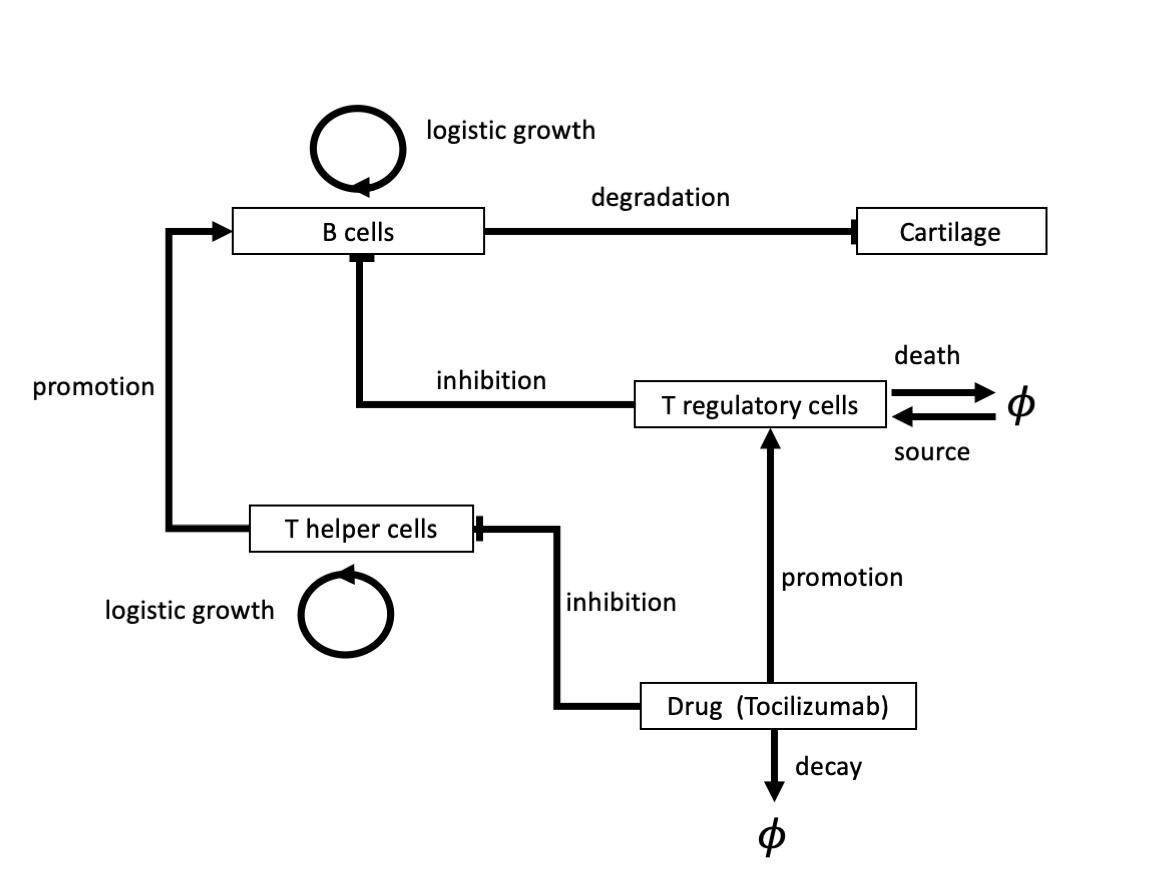}
\caption{ Schematic describing the chemicals, cells and processes included in model described in~\cite{odisharia2017mathematical}. The~models considers five variables; B cells, T helper cells, T regulatory cells, cartilage and the drug TCZ, along with the interactions between these components. Note that, here $\phi$ represents objects outwith or removed from the system.}
\label{Odishara}
\end{figure}

An even larger system of ODEs was proposed by Rao~et~al.~\cite{rao2016mathematical} to describe the circadian dynamics involved in the progression of rheumatoid arthritis. The~hypothalamic-pituitary-adrenal (HPA) axis, which is involved in the regulation of the immune system, is also though to be an important regulator of RA through its modulation of the secretion of pro-inflammatory cytokines. In~\cite{rao2016mathematical}, the authors took a systems approach to identify the types of regulations that have to be included in a signalling network, for~the interactions between mediators of the HPA axis and pro-inflammatory cytokines. This~allowed the results to qualitatively match the observed circadian pathophysiological features of experimental mouse models of arthritis. To this end, they developed 34 coupled ODEs for biological interactions that take place across three compartments: the HPA axis compartment (focused on the circadian secretion of corticosterone; 15 ODEs), a peripheral compartment (focused on the downstream effects of corticosterone on pro-inflammatory cytokines; 18 ODEs), and~a disease endpoint compartment (focused on paw edema, which characterises the severity of experimental arthritis; 1 ODE); see Figure~\ref{Rao2016_Figure}. The~variables modelled in~\cite{rao2016mathematical} include hormones (e.g., corticotrophin-releasing hormone, adrenocorticotropic hormone), receptors (glucocorticoid receptor), and~cytokines (TNF-$\alpha$, IL-6, IL-1$\beta$). The~parameters, of which there were approximately 50, were estimated or taken from published mouse data. The~model was used to investigate two different hypothetical mechanisms by which a tolerance mediator might act on the HPA axis to reduce the secretion of corticosterone, and~impact the evolution of RA.

\begin{figure}
\centering
\includegraphics[width=.98\textwidth]{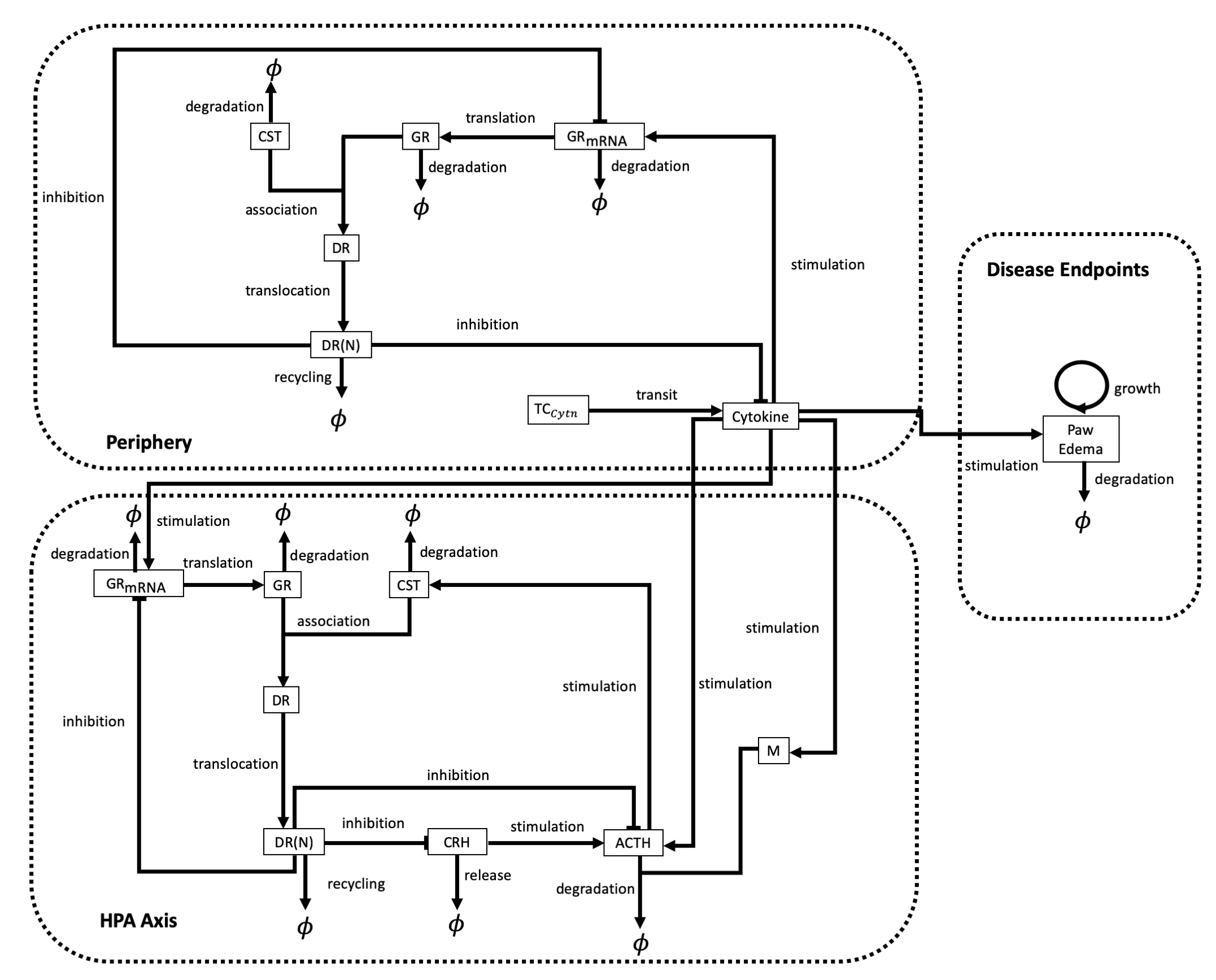}
\caption{Schematic describing the chemicals and processes included in model described in~\cite{rao2016mathematical}. The~model consists of multiple compartments, and~considers the interactions between components within the same compartment and across the compartments. The~abbreviations used are; corticosterone (CST), adrenocorticotropic hormone (ACTH), corticotropin-releasing hormone (CRH), free glucocorticoid receptor (GR), glucocorticoid receptor mRNA (GRmRNA), cytoplasmic corticosterone-bound GR (DR), nuclear corticosterone-bound GR (DR(N)), tolerance mediator (M) and cytokine transit compartment (TCCytn). Note that, here $\phi$ represents objects removed from the system.}
\label{Rao2016_Figure}
\end{figure}

Another example of a large scale system of ODEs in RA modelling is the RA PhysioLab platform, which utilises hundreds of ODEs to simulate the inflammatory and erosive processes occurring at the cartilage-pannus junction~\cite{rullmann2005systems}. The~model consists of a synovial compartment, a cartilage compartment and a bone compartmentm in line with the three main areas of RA activity. A~large number of cell types and cytokines are considered. For~example, the~cell types include macrophages, FLSs, T helper cells, endothelial cells, chondrocytes and osteoclasts. Some of the cytokines and other molecules considered in~\cite{rullmann2005systems} include; TNF-$\alpha$, IL-1$\beta$, IL-6, IFN-$\gamma$, GM-CSF, RANKL, MMPs and TIMPs. The~baseline parameters of the model were chosen to simulate an untreated early stage RA patient, with chronic inflammation and progressive cartilage degradation. The~parameter values were calibrated using published data on rheumatoid cells or joints. In~\cite{rullmann2005systems} the authors used this PhysioLab modelling platform to explore the roles of IL-12 and IL-15 targeting therapies, and~predicted that anti-IL-15 therapy will likely be effective in the virtual patient modelled through this platform. A~further benefit of this modelling approach is that the interactions included can occur at different timescales spanning minutes to months, allowing the user to simulate long-term dynamics of disease progression. The~PhysioLab platform approach has also formed the basis for other models in the field of bone modelling and research~\cite{defranoux2005silico}.

While the ODE models discussed above focused on the immunological mechanisms behind RA development and treatment, there is also a large class of ODE models that focus exclusively on the pharmacokinetics and pharmacodynamics of various drugs used to treat RA; see, for~example, \mbox{\cite{Mould1999_PKPD_Clenoliximab_RA,Ng2005_PKPD_RituximabRA,Kimura2012_TheoryBasedPharmacokineticsRA,Kimura2014_PharmacokineticPharmacodynamicRA,Ternant2013_PharmacokineticsInfliximabRA,Liu2013_PKPD_MethotrexateRA,Levi2013_PKPDmodel_Tocilizumab_RA,Namour2015_PKPDmodel_Filgotinib_RA}} and references therein. For~example, \cite{Ternant2013_PharmacokineticsInfliximabRA} used a two-compartment pharmacokinetic ODE model to study the kinetics of the drug infliximab in RA patients, and~how the kinetics are affected by inflammatory activity and methotrexate co-treatment. A~slightly more complex pharmacokinetic and pharmacodynamic model was introduced in~\cite{Kimura2014_PharmacokineticPharmacodynamicRA} to investigate the anti-inflammatory effects of three anti-TNF inhibitors: infliximab, etanercept and adalimumab. The~authors coupled an algebraic equation for the serum concentration of the anti-TNF-$\alpha$ drugs, with ODEs for the time-evolution of TNF-$\alpha$ and of complexes formed when between TNF-$\alpha$ and TNF inhibitors. Moreover, they introduced an additional differential equation for the quantification of the clinical inflammatory response generated by TNF-$\alpha$. Thus, this model connects molecular-level dynamics of TNF-$\alpha$ and TNF inhibitor drugs with the tissue-level inflammation generated by these cytokines. Using clinical data, the~authors showed that different therapeutic dose regimes with TNF inhibitors can explain the fluctuations in the observed clinical responses, and~suggested that one can predict individual clinical efficacy of these inhibitors by measuring the serum concentration of TNF-$\alpha$ before the treatment. We~conclude this section by mentioning that pharmacokinetic models have also been used to study the kinetics of MRI contrast agents (e.g., gadolinium-diethylenetriame pentaacetic acid, or~Gd-DTPA) in the context of MRI imaging for various tissues involved in RA~\cite{Workie2004_PharmacokineticModelMRI}.
\begin{figure}
\centering
\includegraphics[width=0.7\textwidth]{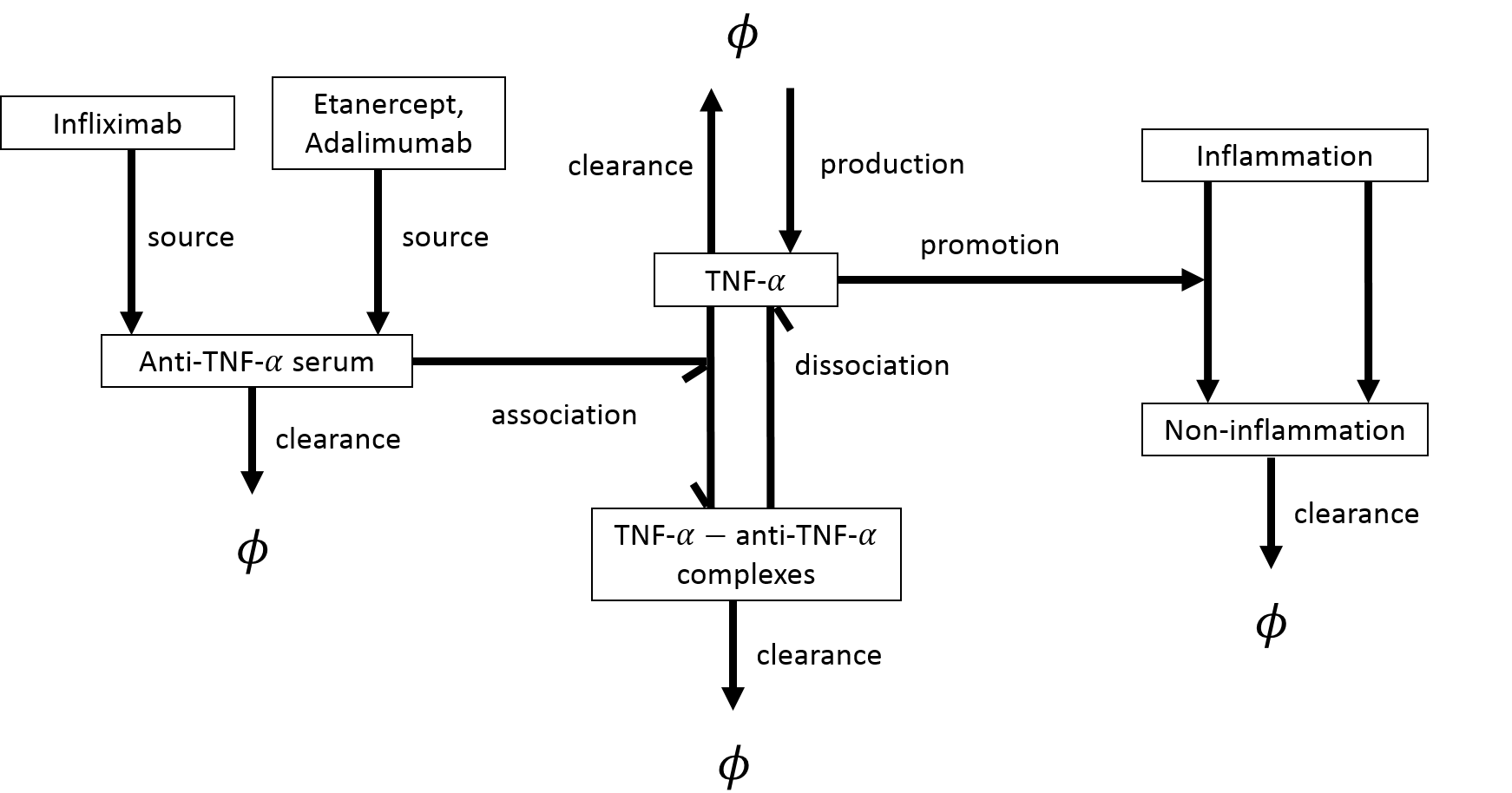}
\caption{Schematic description of the kinetic processes investigated in~\cite{Kimura2014_PharmacokineticPharmacodynamicRA} with the help of a pharmacokinetic and pharmacodynamic model (where ``pharmacokinetics'' refers to drug dynamics through the body, while ``pharmacodynamics'' refers to the body's response to the drugs). Note that, here, $\phi$ represents components outwith the system or components removed from the system.}
\label{Fig_Kimura}
\end{figure}

\subsection{Deterministic Models for Disease Progression and Treatment: PDEs}\label{Sect:PDEs}
 
In the previous subsection, we described ODE models which can only capture the temporal dynamics of a particular system. On the contrary, partial differential equations (PDEs) can be used to consider both spatial and temporal aspects of biological mechanisms. PDEs, and~systems of PDEs, have been used to describe the spatio-temporal dynamics of cells and molecules (cytokines or drugs) within the synovium and the surrounding joints during the pathogenesis and treatment of RA. 

\begin{figure}
\centering
\includegraphics[width=1\textwidth]{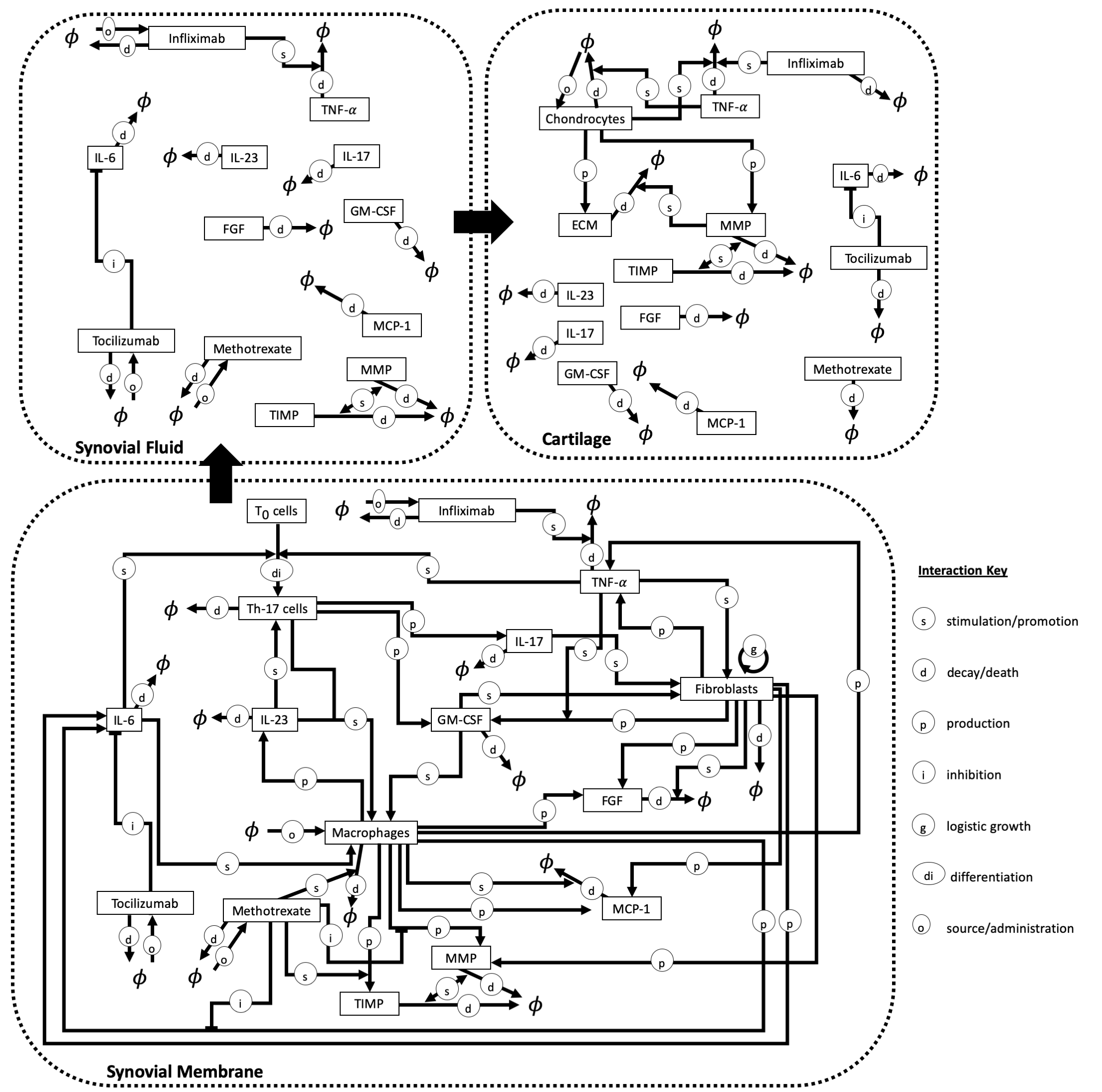}
\caption{Schematic describing the chemicals, cells and processes included in the model described in~\cite{moise2019rheumatoid}. The~model described the interactions between cells, cytokines and drugs in three compartments, the~synovial membrane, the~synovial fluid and the cartilage of joints. In the model, small particles, such as the chemicals, can move between the compartments whereas cells remain within their original compartment. Further to the mechanisms illustrated in the figure the authors additionally consider the diffusion of the cytokines and the movement of cells within each compartment. All abbreviations are those which have been used throughout this review and here $\phi$ represents objects outwith or removed from the system. }
\label{Moise}
\end{figure}

The general structure of these PDEs, for~some abstract variable $U(t,x)$ (which can describe, for example, the density of immune cells or the concentration of cytokines at time $t$ and spatial position~$x$) can be described as follows:
\begin{equation}
\underbrace{\frac{\partial U(t,x)}{\partial t}}_{\text{time-changes in}\;U} = \underbrace{D\frac{\partial^{2} U(t,x)}{\partial x^{2}}}_{\text{diffusion/random movement of}\;U} - \underbrace{B\frac{\partial U(t,x)}{\partial x}}_{\text{advection/directed movement of}\;U}+\underbrace{F(U(t,x))}_{\text{production/decay of}\;U},
\end{equation}
where $D$ is the diffusion coefficient and $B$ is the advection coefficient. The~term $F(U)$ describes the temporal production/decay of variable $U$, as given by the arrows in the diagrams shown in Figures~\ref{Fig_logistic}--\ref{Moise}.

A study that focused on the spatio-temporal dynamics of immune cells, cytokines, therapeutic drugs, and~the spatio-temporal degradation of the cartilage, was introduced by
Moise~et~al.~\cite{moise2019rheumatoid} who developed a system of 17 PDEs (in one spatial dimension) to model the progression of the disease and to evaluate the success of RA drug treatments. They considered a 3-compartment model that incorporated various biological mechanisms occurring within the synovial fluid, synovial membrane and the cartilage of a joint; see Figure~\ref{Moise}. Some of the cell types considered in~\cite{moise2019rheumatoid} include Th17 cells, macrophages, fibroblasts and chondrocytes. Note that the authors chose to not include B cells as they focused on chronic/established RA. Regarding the cartilage, the~authors considered the density of the extracellular matrix (ECM) that could be degraded by MMPs. Furthermore they considered several cytokines and growth factors (e.g., IL-6, IL-17, TNF-$\alpha$, GM-CSF). The~aim of this study, which incorporated space in the dynamics of all cells and molecules, was to evaluate the impact of three RA drugs on RA progression that was quantified by the width of the cartilage layer. The~three drugs were methotrexate (MTX) (which increases the apoptosis of macrophages), infliximab (IFX) (a TNF-$\alpha$ inhibitor), and~tocilizumab (TCZ) (which blocks the IL-6 receptor). Overall, the~model included over 100 parameters that were estimated from various published studies. The~authors presented some numerical simulations for the time-evolution of various cells and cytokines, and~for the degradation of the cartilage as quantified by the movement of the synovial membrane. We~should emphasise, however, that this model only considered the benefits of drugs and not any potential side effects.
 
 \subsection{Stochastic Models}
 \label{Sect:Stochastic}
All of the models we have described have been deterministic, however, in the field of mathematical biology it has become more common to incorporate stochastic terms to capture the randomness of different biological mechanisms. Accounting for stochastic environmental events seems to be particularly important for RA development and evolution~\cite{firestein2017immunopathogenesis,de2017difficult}, as well as for RA treatment~\cite{Eseonu2015_HomingMSCStochastic} (e.g., in~the context of the deterministic or stochastic migration of mesenchymal stem cells during the cartilage repair processes). However, this stochastic approach is not very common in the current mathematical models.

One of the few models that account for stochastic environmental/genetic effects was introduced in~\cite{RobertsThompson2002_StochasticRA}, where the authors considered a simple (1-equation) stochastic mathematical model to describe the age-specific RA incidence rate as a function of the number of random events that occur before the disease manifests. The~authors compared the results of their stochastic model with population data from the Australian Bureau of Statistics, and~concluded that only a small number of events (e.g., environmental or genetic random events) have to occur in a predisposed population to allow for the emergence of the disease. A~slightly different approach was considered in~\cite{helliwell2000joint}, where the authors aim to predict the probability of symmetry of joint involvement in early and late RA. Similarly, in~\cite{wick2004relationship} the authors consider the relationship between radiological progression and inflammation, which they find to be patient specific. 

\subsection{Probabilistic Cost-Effectiveness Models for RA Treatment Strategies}\label{Sect:CostEffectiveness}

An alternative research area that has arisen over the past two decades focuses on the development of cost-effectiveness models for different RA treatments and combinations or sequences of treatments. These health economic models have become common tools for decision making in regard to different RA therapies, as they take into account the cost of the therapies (i.e., costs of the drugs), their effectiveness (i.e., some drugs are more effective than others), as well as potential complications. We~have seen in Section~\ref{Sect:Treatment} that there exists already a large number of therapeutic drugs for RA, and~more new compounds are currently being developed~\cite{Kalden2016_EmergingTherapiesRA}. However, there are differences between these drugs in terms of their efficacy as well as in terms of their costs~\cite{Nurmohamed2005_EfficacyTolerabilityCostDrugsRA}. For~example, in~regards to efficacy, DMARDs and biologics can lead to clinical disease control or remission, while NSAIDs can inhibit rapidly the local inflammatory symptoms but have almost no lasting effects on the systemic aspects of RA~\cite{Geyer2010_RationaleDifferentTherapiesRA}.

In general, biological agents are more expensive than the conventional DMARD approaches, but lead to an increased quality of life. Therefore, cost-effectiveness approaches have become recognised as essential to allocate healthcare resources~\cite{Scholz2014_ModelRA} or to design patient-oriented treatment plans~\cite{Geyer2010_RationaleDifferentTherapiesRA}. There are various decision models in the literature ranging from Markov-chain models, to decision trees, discrete event simulations, or~individual sampling methods~\cite{Scholz2014_ModelRA}. In the following we focus briefly on some of the Markov-chain models, since these are the most commonly used models~\cite{Scholz2014_ModelRA}. 

\begin{figure}
\centering
\includegraphics[width=0.7\textwidth]{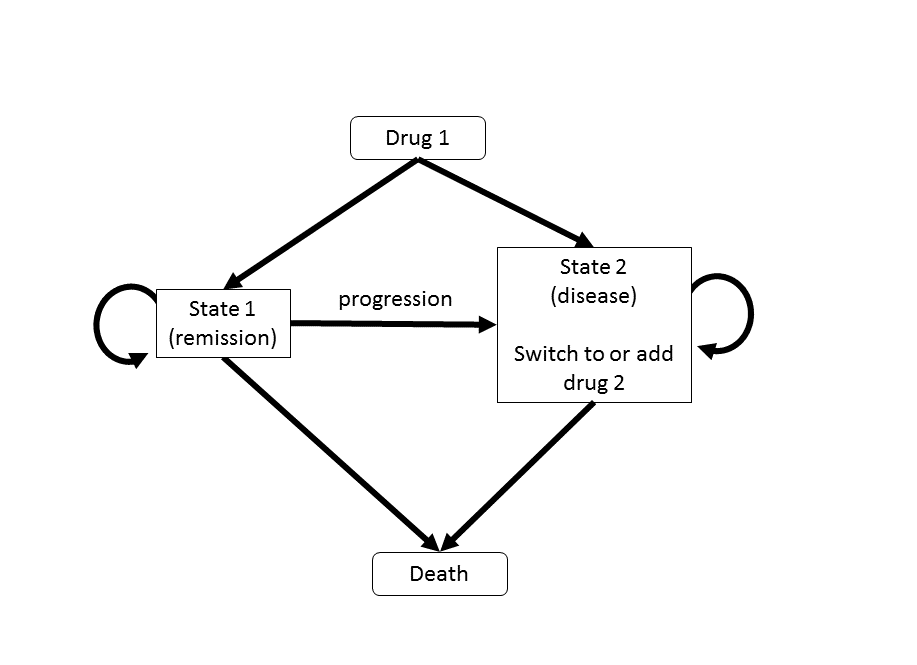}
\caption{Schematic description of the structure of a Markov model with 2 states (``remission'' and ``disease''), under a treatment based on two drugs. The~simulations start with ``drug 1'', and~if the patients do not respond to the treatment then they are switched to ``drug 2'' (or ``drug 2'' is introduced in combination with ``drug 1''). }
\label{MarkovModels}
\end{figure}

The Markov-chain models (see the brief description of such a model in Figure~\ref{MarkovModels}) are discrete-time probabilistic models, where the probability of being in a given disease state (e.g., ``remission'', ``low disease activity'', ``moderate disease activity'' or ''high disease activity''~\cite{Schipper2011_TreatmentStrategiesRA}, or~even ``death''~\cite{Spalding2006_CostEffectivenessRA_TNFa}) at a given time depends only on the probability distribution over all states in the previous time step, and~the transitions rates linking these states. Some of these models for RA focus on single drugs used on multiple patient cohorts (e.g., infliximab ~\cite{Lekander2013_CostEffectivenessRAinfliximab}), while other models investigate the effectiveness of single drugs versus combinations of drugs (e.g., methotrexate vs. methotrexate+anti-TNF~\cite{Schipper2011_TreatmentStrategiesRA}, or~etanercept vs. methotrexate vs. methotrexate+etanercept~\cite{Kobelt2005_CostEffectivenessRA}). In this case, if the treatment with a specific drug is not effective (e.g., patients might leave the ``remission'' state), then the treatment can be changed and the patients be put on a different drug or combinations of drugs. The~transitions between states occur at specified intervals, i.e., ``cycles'', which can vary between studies; e.g., 1-year cycles in~\cite{Kobelt2005_CostEffectivenessRA}, 6-months cycles in~\cite{Kobelt2011_CostEffectiveness}, or~3-months cycles in~\cite{Maetzel2003_CostEffectiveness}. Moreover, these transition probabilities are calculated based on observed transitions in clinical trials. 

Since the literature of cost-effectiveness simulation Markov models has grown exponentially over the past years, we will not review more models here, but rather refer the reader to the studies in~\cite{Scholz2014_ModelRA,Schipper2011_TreatmentStrategiesRA,Spalding2006_CostEffectivenessRA_TNFa,Lekander2013_CostEffectivenessRAinfliximab,Kobelt2005_CostEffectivenessRA,Zhang2018_CostEffectivenessRA,Kostic2017_CostEffectivenessRA_Serbia,Brennan2003_CostEffectivenessRA_UK,Jalal2016_CosteEffectivenessRA_TripleTher,Kobelt2011_CostEffectiveness,Maetzel2003_CostEffectiveness,Alemao2018_RAconceptualModelHealthAssess} and the references therein. However, in~Section~\ref{Sect:Conclusion}, we will return to these types of models and discuss their importance on personalising RA treatments for different patients in different countries, which might impact also the deterministic mathematical models used to understand the biological mechanisms behind the evolution of the disease.

\begin{Remark}
Since these stochastic/probabilistic models are mostly computational (with probabilistic occurrence or transition rates being considered at different time steps in the numerical simulations), we cannot write down mathematical equations. We~emphasise here that classical stochastic (ordinary and partial) differential equations~\cite{Oksendal_StochasticDiffEq,Dalang_StochasticPDEs} have not been developed yet in the context of RA.
\end{Remark}

\subsection{Parameter Estimation and Availability of Data}\label{Sect:Data} 

For a mathematical model to be biologically relevant, most (if not all) of the parameters must be chosen to be consistent with biological data. Optimally, the~model will only require estimation for a small number of the parameters used. Therefore, the~complexity of a mathematical model depends heavily of the availability of biological data to support parameter choices. This~biological data can come from in vitro experiments, in~vivo animal models or human studies, e.g., clinical trials. These different types of studies each have their own benefits and drawbacks. For~example, in~vitro experiments are cheap and allow for control over the mechanisms studied. However, in~vitro experiments do not replicate the full biological dynamics and chronic effects cannot be tested. On the other hand, in~vivo models (animal models) do allow for a more biologically relevant environment to be investigated and new treatment approaches to be tested. However, with in vivo studies there is less control over the background environment, higher cost/time scales and ethical considerations that must be accounted for. Furthermore, in~the context of RA, autoimmune conditions that are present in humans do not develop in the same way as in animals, and~have to be induced therapeutically with a short disease lifetime~\cite{mina2015basics}. Finally, clinical trials or human studies give the optimal biologically relevant setting, however, the~information that can be obtained from patients is limited. For~example, in~RA patients blood tests, X-rays and disease activity scores can be considered, however, more in depth information such as cell interactions and live cell tracking cannot be achieved. Therefore in depth patient specific data can be hard to obtain. Generally, in~the development of new RA treatments all three approaches to experimental modelling can be utilised. Initially, biological mechanisms to target within RA must be firstly identified. To find these targets in vitro~\cite{peck2018establishment} and in vivo models can be used, with in vivo models then used to further validate these targets~\cite{mina2015basics,sardar2016old}. After initial drug development, in~vivo studies can be used to assess the pharmacodynamic and pharmacokinetics of the proposed drug in response to RA~\cite{mina2015basics}. Finally, after pre-clinical testing the drug then can go forward to clinical trials~\cite{kirkham2013outcome}. The~data from clinical trials can additionally be compared with previous, similar, trials to investigate correlations between patients and assess the efficacy of the new treatment~\cite{campbell2010risk,pawar2019risk,adalimumabref,etanerceptref,ifxref,mtxref,singh2015risk}.

Returning to the various mathematical models discussed in this review, we make the following observations:
\begin{itemize}
\item The mathematical models used to describe the pharmacokinetics and pharmacodynamics of RA drugs have all been parametrised using patients data from clinical trials, as well as using different laboratory analyses; see~\cite{Ternant2013_PharmacokineticsInfliximabRA,Kimura2012_TheoryBasedPharmacokineticsRA,Levi2013_PKPDmodel_Tocilizumab_RA,Namour2015_PKPDmodel_Filgotinib_RA}. Similarly, the~cost-effectiveness models for various RA therapies have all been parametrised using patients data from clinical trials; see~\cite{Maetzel2003_CostEffectiveness,Kobelt2005_CostEffectivenessRA,Kobelt2011_CostEffectiveness}. Given the large number of clinical trials on RA (e.g., there are currently more that 2,300 studies on RA listed on the website ``ClinicalTrials.gov''), this might explain the very fast development of these two classes of mathematical models (i.e., pharmacokinetic/pharmacodynamics models, and~cost-effectiveness models) over the last two decades. 
\item From those reviewed, the~only deterministic models for RA evolution that have been parametrised exclusively with patients data were the ODE models for the radiographic progression of RA (where the data was obtained following the radiographic examination of patients' joints); see~\cite{Graudal1998_RadiographicProgressRA,Witten2000_OneODE_ProgressionRA}. 
\item The majority of mathematical models used to describe disease progression used a mixture of in vitro, in~vivo and human data, with additional unknown parameters being estimated, as in some of the models we have described earlier in this work~\cite{jit2004tnf,moise2019rheumatoid}. In these models, data such as cytokine decay rates comes from non-RA specific in vitro studies, while cytokine concentrations are taken from RA specific studies~\cite{moise2019rheumatoid}. Combining different types of data to parametrise the models can lead to uncertainty in the parameter values. This~uncertainty is increased by variability in the data from different patients (and cohorts of patients), or~by variability in the experimental set-ups. Therefore, using a (reasonable) range of values for each parameter and undertaking sensitivity analysis may prove beneficial.
\end{itemize}

\section{Conclusions}\label{Sect:Conclusion}

In this review we have summarised some quantitative predictive modelling approaches developed over the last 20--30 years to understand the complex autoimmune responses involved in the development and evolution of rheumatoid arthritis. To understand the biological aspects investigated by these modelling approaches, we started by focusing first on the biology of this disease and the current therapies aimed at controlling it. We~then discussed briefly different types of mathematical models introduced to describe different aspects in the development and evolution of RA. To this end, we focused not only on the simplicity-vs.-complexity of these models, but also on deterministic-vs.-stochastic processes investigated, as well as the scale at which these models were derived (i.e., molecular-, cell- and tissue/joint-level scales). We~finished by mentioning a particular class of cost-effectiveness probabilistic (Markov chain simulation) models developed to quantify the decisions behind various therapeutic approaches to RA. All these types of deterministic and probabilistic mathematical models have been summarised in Table~\ref{TableModels}. 

We need to emphasise that despite of the numerous models for autoimmunity that exist in the mathematical literature, we could not find too many models aimed specifically at RA (in the context of biological mechanisms for disease progression). While some of these published models for autoimmune diseases could be applied to RA, it~is known that various autoimmune diseases might behave in different ways, as discussed in~\cite{jit2004tnf} in the context of rheumatoid arthritis (RA) versus systemic inflammatory response syndrome (SIRS). Therefore, to investigate the different complex immune aspects involved in the evolution of RA, as well as the genetic/environmental factors that could trigger this disease, it~is important that new deterministic and stochastic models are being derived in the future (at single and multiple scales). 

\subsection*{Future Predictive Modelling Approaches}
We believe that the next two-three decades will see the development of new mathematical models for RA, aimed at proposing new hypotheses for the immunological mechanisms behind the progression of this disease. With the continuous development of experimental approaches that would generate more immune-related experimental/clinical data, many of these models will also be validated experimentally, for~more biological realism and impact on therapeutic decisions. We~have seen in Section~\ref{Sect:CostEffectiveness} that the cost-effectiveness models incorporate some descriptive data on the evolution of the disease (in addition to various other quantitative data on the cost of the therapies, and~qualitative data on the side effects of these drugs). As discussed in Section~\ref{Sect:Data}, many deterministic mathematical models for RA are not fully parametrised with experimental/clinical data, and~while they are very useful for qualitative predictions on model dynamics, they cannot be used for quantitative predictions. 

\emph{Cost-effectiveness models.} It would be interesting and potentially useful to combine various probabilistic cost-effectiveness Markov-chain models with deterministic (ODE or PDE) models for the evolution of the disease under various treatments and combinations of treatments that have been determined to be more effective. For~example, \cite{Kostic2017_CostEffectivenessRA_Serbia} suggested that for Serbian RA patients, treatment with methotrexate alone is more cost effective than a combination treatment with etanercept+methotrexate. This~is in contrast to an earlier study on UK patients~\cite{Brennan2003_CostEffectivenessRA_UK}, where the etanercept was suggested as being more cost effective compared to classical DMARD agents. These cost-effectiveness models likely influence decisions on treatment approaches in different countries, and~by taking them into consideration when developing new deterministic/stochastic models for the evolution of the RA disease, they could help us investigating personalised treatment approaches for patients in different countries.

\emph{Side-effects.} As mentioned in Section~\ref{Sect:Treatment} many of the current drugs used in RA treatment can have side effects, which can occasionally be severe or even fatal. With the exception of some cost- effectiveness models, the~majority of the mathematical models that we have described in this review did not investigate the potential negatives (side effects) of the drugs they were investigating. Therefore, it~may be of potential benefit to use mathematical modelling approaches, e.g., methods similar to those we have reviewed, to consider the potential adverse effects of these drugs.

\emph{Stochastic individual-based models, multi-scale models and hybrid models.} The development of stochastic models is particularly important for a better understanding of RA, where the aetiology of the disease is still not fully understood. Furthermore, various experimental works highlight the influence of stochastic environmental events and heterogeneity within the development and evolution of the disease. Individual-based models describe each cell as a single agent that can act independently of other cells. Within these models, the~actions of each cell can be probability-based rather than deterministic, allowing for the inclusion of stochasticity within the system. This~approach to modelling has been used extensively in mathematical modelling within ecology~\cite{Grimm2005}, biology~\cite{van2015simulating,An2009} and, more specifically, the~modelling of other diseases, such as cancer~\cite{metzcar2019review}. We~believe that such stochastic models could be developed to describe random processes occurring within rheumatoid arthritis across all scale levels: molecular, cell, and~tissue/joint scales. These models could also connect the three compartment levels depicted in Figure~\ref{Fig_Multiscale}, to generate new multi-scale mathematical models. Multi-scale methods have previously been used to describe biological and medical phenomena~\cite{martins2010multiscale}. These multi-scale methods may use hybrid modelling approaches, whereby, deterministic ODEs (and PDEs) and stochastic methods may be used to model mechanisms at different spatial scales, feeding information to the other scales and contributing to the complete model. The~development of such models has been successful in the description of the progression and treatment of cancer~\cite{Macklin2010,rejniak2011hybrid,Lowengrub2009}. For~example, Powathil~et~al.~\cite{powathil2015systems} developed a multi-scale hybrid model to describe cancer progression to predict the effects of radiotherapy and chemotherapy. In their model, ODEs were used to describe the cell cycle dynamics, PDEs were used to model oxygen and treatment effects which both fed into an agent-based model that described the cellular level interactions. In a similar way, in~the context of rheumatoid arthritis, using multi-scale hybrid modelling approaches may be valuable in modelling disease progression and predicting the success of RA treatment.

\section*{Abbreviations}
{The following abbreviations are used multiple times throughout this manuscript:\\

\noindent 
\begin{tabular}{@{}ll}
ACPA & Anti-citrullinated protein antibody\\
CRP & C-reactive protein\\
DMARD & Disease modifying anti-rheumatic drug\\
ECM & Extracellular matrix\\
ESR & Erthrocyte sedimentation rate\\
FLS & Fibroblast-like synoviocyte\\
GM-CSF & Granulocyte-macrophage colony-stimulating factor\\
IFN-$\gamma$ & Interferon - $\gamma$ \\
IFX & Infliximab\\
IL- 'x'& Interleukin - 'x'\\
MMP & Matrix metalloproteinase\\
MTX & Methotrexate\\
NF-$\kappa$B &Nuclear factor-$\kappa$ B \\
NSAID & Non-steroidal anti-inflammatory drug\\
ODE & Ordinary differential equation\\
PDE & Partial differential equation\\
RA& Rheumatoid arthritis\\
RANK(L) & Receptor activator of NF-$\kappa$B (ligand)\\
TCZ & Tocilzumab\\
TIMP & Tissue inhibitor of metalloproteinase\\
TNF-$\alpha$ & Tumour necrosis factor - $\alpha$\\
TNFi & TNF-$\alpha$ inhibitor
\end{tabular}}

\bibliographystyle{plain}
\bibliography{references}

\end{document}